\documentclass[journal,10pt]{IEEEtran}
\usepackage{amsmath,amsfonts}
\usepackage{algorithmic}
\usepackage{algorithm}
\usepackage{array}
\usepackage[caption=false,font=normalsize,labelfont=sf,textfont=sf]{subfig}
\usepackage{textcomp}
\usepackage{stfloats}
\usepackage{url}
\usepackage{verbatim}
\usepackage{graphicx}
\usepackage{cite}
\usepackage{setspace}
\usepackage{color}
\begin{document}

\title{Ubiquitous Integrated Sensing and Communications for Massive MIMO LEO Satellite Systems}

\author{Li You, Yongxiang Zhu, Xiaoyu Qiang, Christos~G. Tsinos, Wenjin Wang, Xiqi Gao, and~Bj\"{o}rn~Ottersten
\thanks{Li You, Yongxiang Zhu, Xiaoyu Qiang, Wenjin Wang, and Xiqi Gao are with the National Mobile Communications Research Laboratory, Southeast University, Nanjing 210096, China, and also with the Purple Mountain Laboratories, Nanjing 211100, China (e-mail: lyou@seu.edu.cn; zhuyx@seu.edu.cn; xyqiang@seu.edu.cn; wangwj@seu.edu.cn; xqgao@seu.edu.cn).}
\thanks{Christos G. Tsinos is with the Department of Digital Industry Technologies, National and Kapodistrian University of Athens, 34400, Evripus Campus, Greece (e-mail: ctsinos@uoa.gr).}
\thanks{Bj\"{o}rn Ottersten is with the University of Luxembourg, Luxembourg City 2721, Luxembourg (e-mail: bjorn.ottersten@uni.lu).}}

\markboth{}%
{Shell \MakeLowercase{\textit{et al.}}: A Sample Article Using IEEEtran.cls for IEEE Journals}

\maketitle

\begin{abstract}
The next sixth generation (6G) networks are envisioned to integrate sensing and communications in a single system, thus greatly improving spectrum utilization and reducing hardware costs. Low earth orbit (LEO) satellite communications combined with massive multiple-input multiple-output (MIMO) technology holds significant promise in offering ubiquitous and seamless connectivity with high data rates. Existing integrated sensing and communications (ISAC) studies mainly focus on terrestrial systems, while operating ISAC in massive MIMO LEO satellite systems is promising to provide high-capacity communication and flexible sensing ubiquitously. In this paper, we first give an overview of LEO satellite systems and ISAC and consider adopting ISAC in the massive MIMO LEO satellite systems. Then, the recent research advances are presented. A discussion on related challenges and key enabling technologies follows. Finally, we point out some open issues and promising research directions.
\end{abstract}

\section{Introduction}
\IEEEPARstart{A}{long} with the commercial deployment of the fifth generation networks, the sixth generation (6G) networks are gradually evolving from a vision into concrete designs. 6G networks are envisaged to support lower latency, higher data rates, and more diverse user terminals (UTs), which can be applied to emerging new applications, including sustainable cities, autonomous driving, mobile health, etc. Moreover, it is anticipated that 6G networks will furnish pervasive wireless intelligent services for UTs through wireless connectivity anywhere on earth\cite{chen2020}. 

As a critical part of the space-air-ground-sea integrated network (SAGSIN), satellite communications play a vital part in offering global coverage. Terrestrial networks cover only about 6$\%$ of the whole earth surface\cite{chen2020}. In some environmentally harsh areas, e.g., the poles, oceans, air, and space, it is unlikely to provide communications services by terrestrial networks. Comparatively, satellite communications have several advantages, such as broad coverage, flexible deployment, little interference from terrestrial networks, etc. Low earth orbit (LEO) satellites, with orbits situated at altitudes 200 -- 2000 km, are relatively lower than the geostationary earth orbit (GEO) and medium earth orbit (MEO) ones, leading to lower latency, path loss, and production costs. LEO satellite communications have seen new developments in recent years. Until now, several LEO satellite communications projects have emerged, including Iridium, Globalstar, OneWeb, Starlink, Telesat, and Hongyun\cite{AST}. Apart from that, LEO satellite synthetic aperture radar has found extensive applications for earth remote sensing for more than 30 years. It is an active remote sensor working in the microwave band, and its azimuth resolution can be improved by the Doppler effect arising from the movement of LEO satellites\cite{SDR}. Massive multiple-input multiple-output (MIMO) technology has been widely used in terrestrial networks to serve dozens of UTs over the same time-frequency resources via base stations (BSs) equipped with very large numbers of antennas. Extending massive MIMO technology to LEO satellite communication systems enables the latter to fully utilize the available degrees of freedom in both temporal and spatial domains, significantly improve spectral efficiency, and enhance robustness to interference\cite{you2022aug,you2022oct}.

Sensing functionality is envisaged to be integrated into 6G networks, such that it becomes a native capability that offers various sensing services to UTs, e.g., localization, recognition, and imaging. On the one hand, with the continuous evolution of the wireless communication, there will be an overlap with the traditional sensing frequency bands and communication frequency bands\cite{isacoverview}. It is urgent to realize frequency reuse between sensing and communications. On the other hand, there are more and more similarities between sensing and communications in system design, signal processing, and data processing, as they operate to reach the mmWave band\cite{isacoverview}. ISAC is put forward as a way to enhance the efficiency of valuable spectrum resource utilization and reduce hardware costs. Besides, scalable trade-offs, synergies, and mutual benefits can be realized by co-designing the two functionalities.

Existing ISAC studies mainly focus on terrestrial networks, which can only offer sensing and communications services in a fixed and limited area. By exploiting the ubiquitously global coverage capability of LEO satellite systems, the massive MIMO LEO satellite ISAC systems have great potential to provide ubiquitous and seamless coverage, enhancing communication capability and sensing accuracy. Nonetheless, owing to the distinct attributes inherent in LEO satellite systems, it is not straightforward to directly adopt the previous ISAC works for terrestrial networks to LEO satellite systems. Such a novel space ISAC paradigm encounters some new difficulties. First, LEO satellites have limited payload capabilities, which largely limits the size and weight of sensing and communicaiton hardware. Second, the non-uniform traffic distribution over the coverage area results in a mismatch between the offered and requested resources. In addition, the large Doppler shifts and high propagation delay caused by the long distances between the UTs/targets and the LEO satellites, as well as their mobility, will degrade sensing and communications.

In this paper, we consider adopting ISAC in massive MIMO LEO satellite systems for realizing ubiquitously global higher-capacity communications and more flexible sensing. The rest of the paper is organized as follows. Sec. \uppercase\expandafter{\romannumeral2} offers an overview of LEO satellite systems, ISAC, and the considered massive MIMO LEO satellite ISAC systems. The recent research advances, including multi-satellite-enabled ISAC and reconfigurable intelligence surface (RIS)-assisted satellites for Internet of Things (IoT) networks, are presented in Sec. \uppercase\expandafter{\romannumeral3}. In Sec. \uppercase\expandafter{\romannumeral4}, we identify some challenges and key enabling technologies. In Sec. \uppercase\expandafter{\romannumeral5}, several open issues and promising research directions are discussed. Finally, conclusion is given in Sec. \uppercase\expandafter{\romannumeral6}.
\section{Overview of LEO Satellite Communication Systems, ISAC, and Massive MIMO LEO Satellite ISAC Systems}
In this section, we highlight several characteristics of LEO satellite communication systems and present an overview of the ISAC technology. Then, we present the application of ISAC in massive MIMO LEO satellite systems.
\subsection{LEO Satellite Communication Systems}
LEO satellite communication systems can provide ubiquitous and seamless connectivity to areas with no or inadequate Internet access, showing great potential for complementing and extending terrestrial networks. The key characteristics of LEO satellite communication systems can be summarized as follows.
\begin{itemize}
	\item{Compared with terrestrial systems, LEO satellite systems have longer communication distances and broader coverage. Besides, LEO satellite systems can support the UTs/targets with moving speed beyond Mach level, which is scarcely possible for terrestrial systems.}
	\item{The establishment of LEO satellite links is not restricted by geographical conditions, wheater in urban areas or remote regions. In addition, dense and flexible networking can be easily achieved in LEO satellite systems.}
	\item{The channel characteristics of LEO satellite systems are quite different from those of terrestrial systems in several respects. For instance, the propagation delay exhibits a much larger value than terrestrial systems. When the LEO satellite is at an altitude of 1000 km, and with UT'selevation angle set at $45^{\circ}$, the round-trip propagation delay can reach the value of 17.7 ms\cite{architecture}.}
\end{itemize}
\begin{figure}[!t]
	\centering
	\includegraphics[width=3.5in]{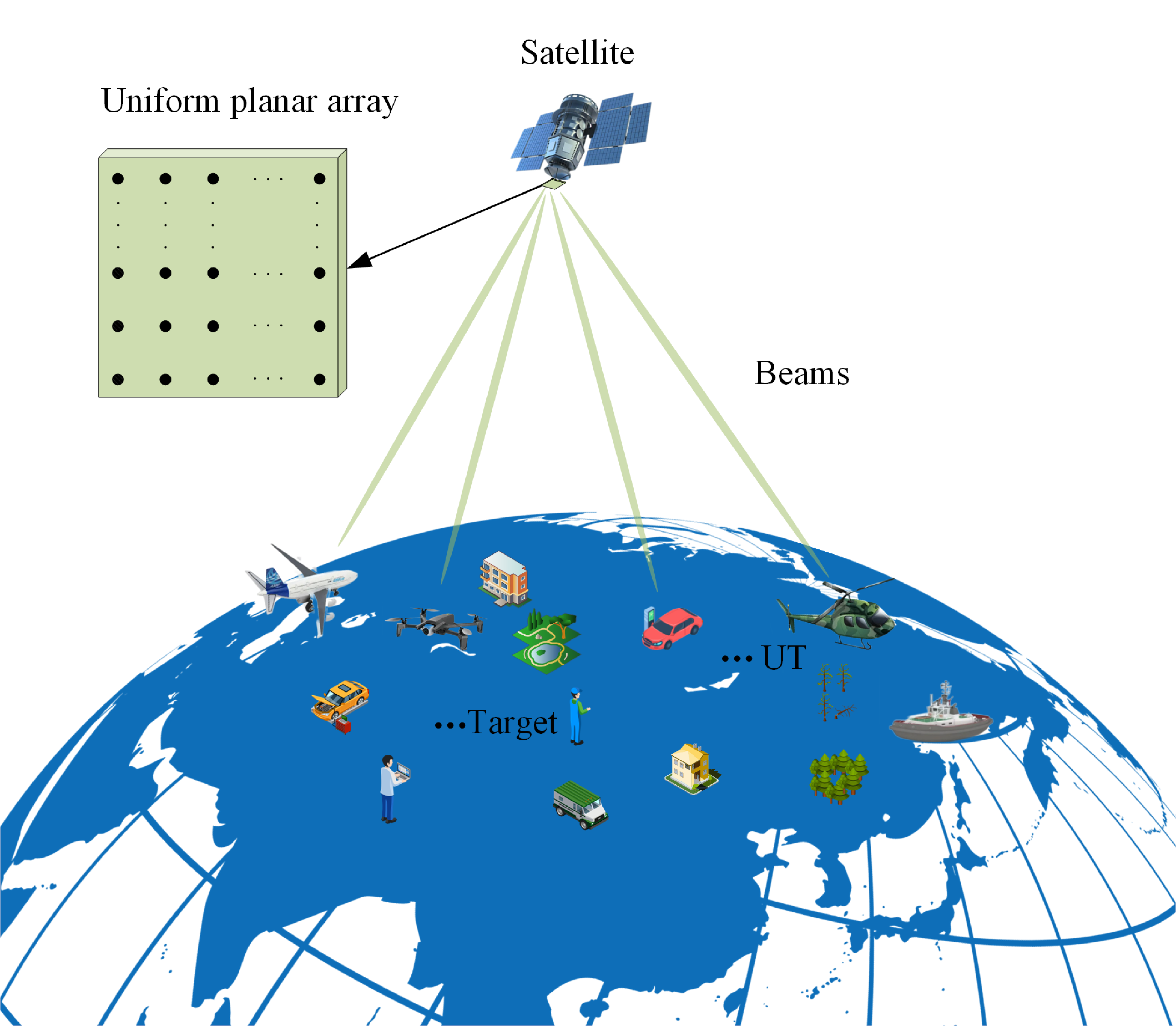}
	\caption{Illustration of a massive MIMO LEO satellite ISAC system setup.}
	\label{fig1}
\end{figure}
\begin{figure*}[htbp]
	\centering
	\subfloat[]{\includegraphics[width=0.33\textwidth]{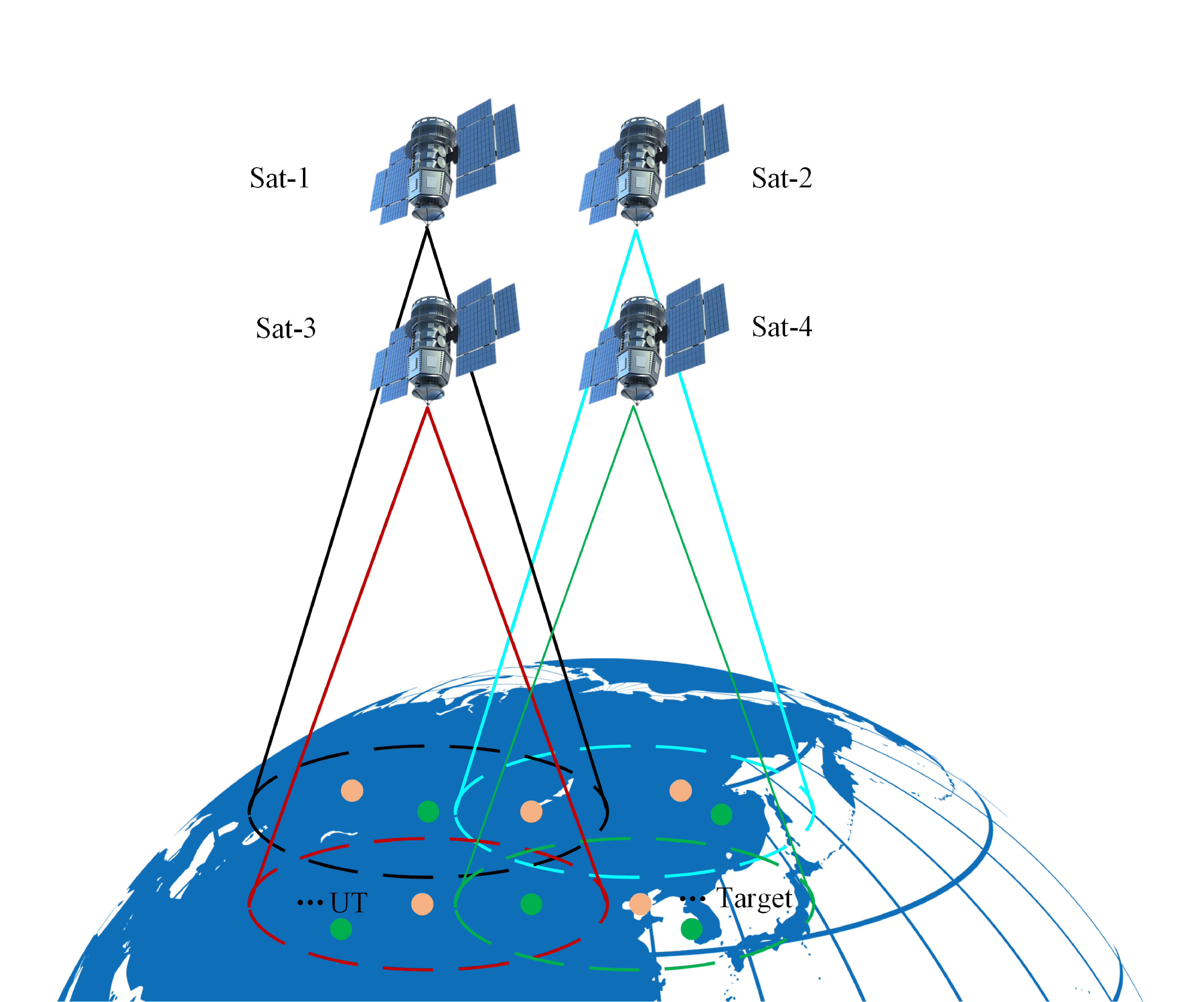}\label{}}
	\hfill
	\subfloat[]{\includegraphics[width=0.33\textwidth]{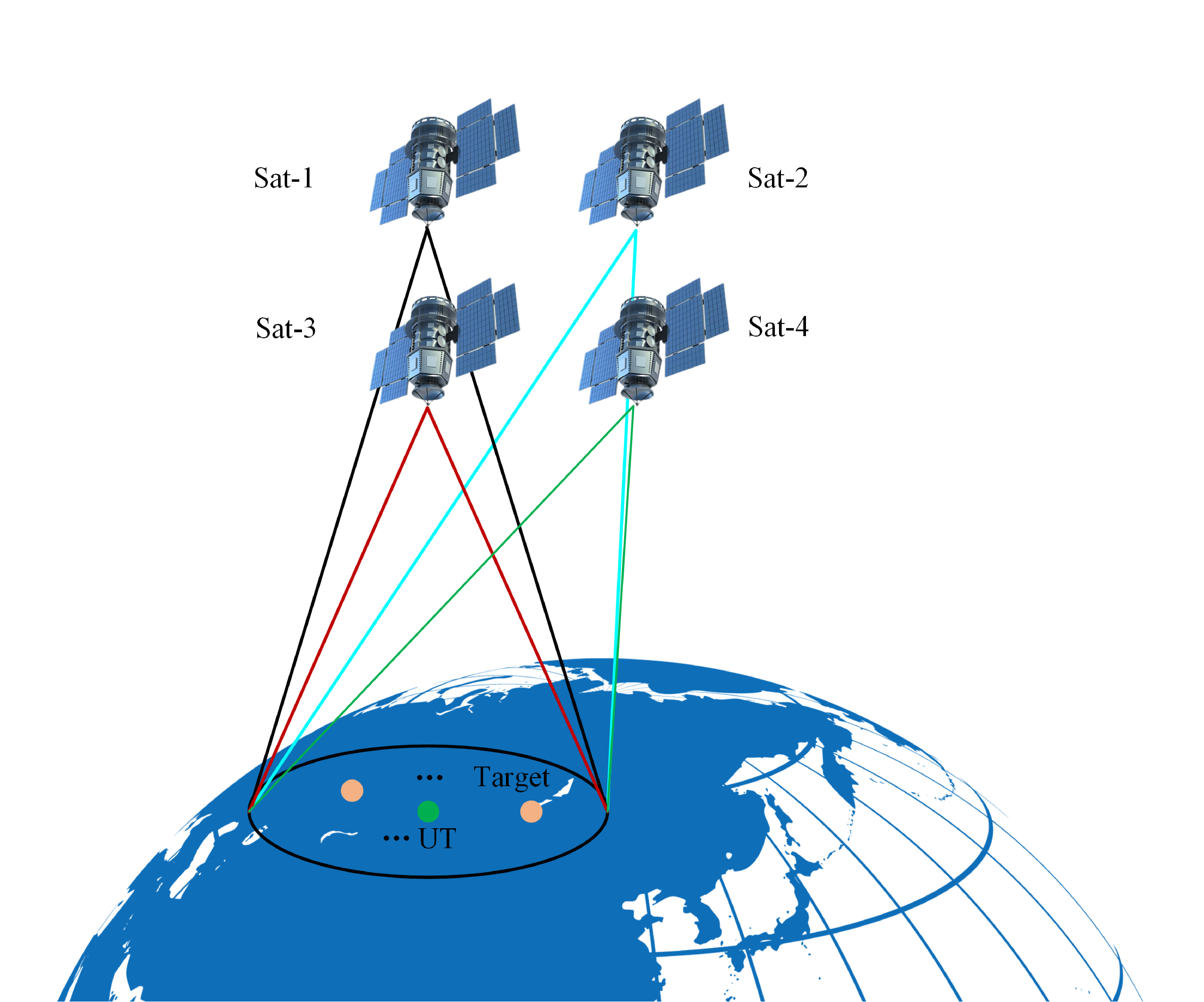}\label{}}
	\hfill
	\subfloat[]{\includegraphics[width=0.33\textwidth]{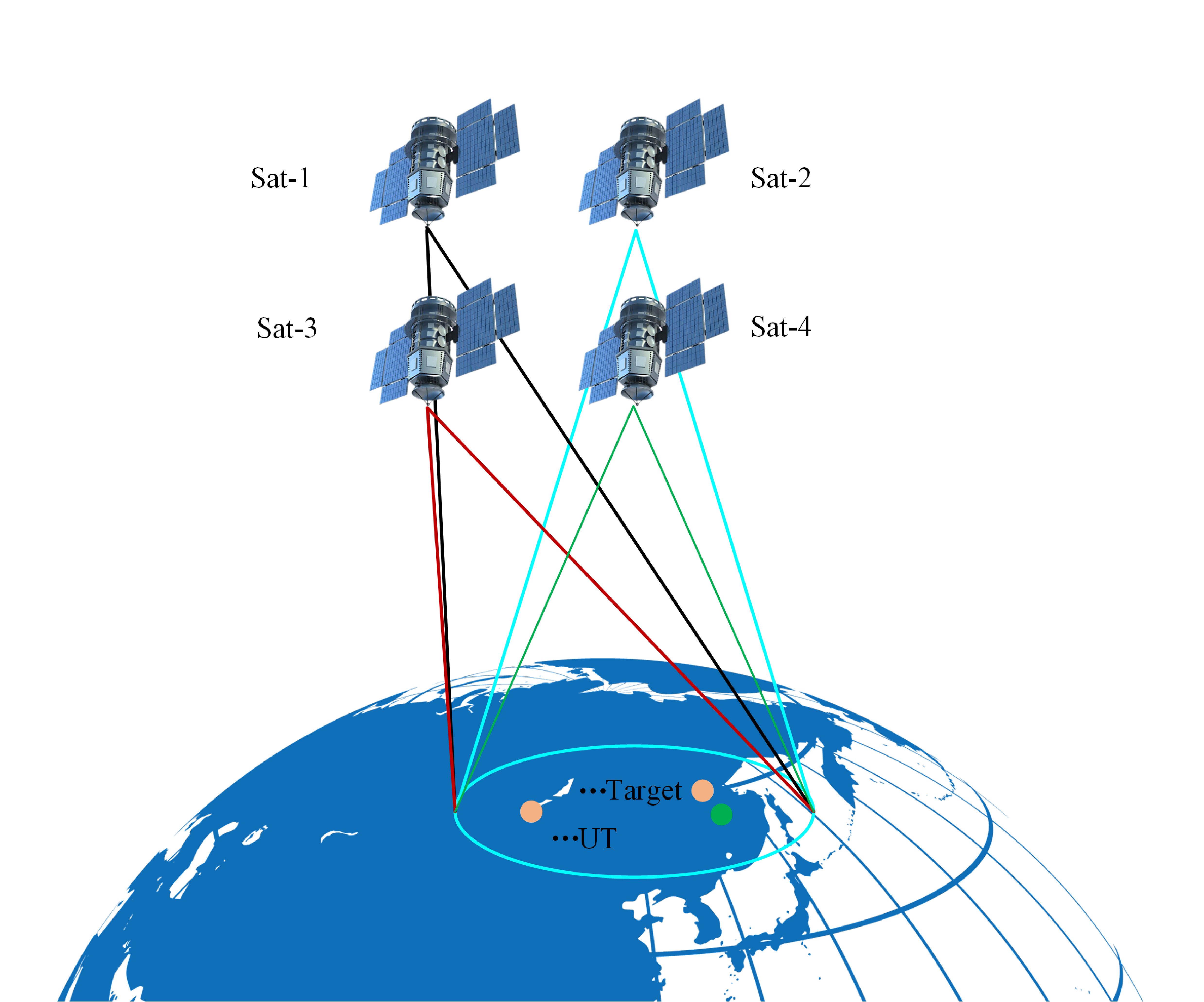}\label{}}
	\caption{(a) Communication beam coverage. (b) Localization beam coverage for Sat-1. (c) Localization beam coverage for Sat-2.}
	\label{fig2}
\end{figure*}
\subsection{Overview of ISAC}
Sensing and communications process information in different ways. As for sensing, it collects information about the sensed objects, while communications focus on how to transfer information in consideration of interference and noise. The early form of ISAC is called radar-communication coexistence (RCC), similar to cognitive radio, wherein radar and communication systems coexist within a shared frequency band, representing a loose integration\cite{isacoverview,Dokhanchi2019}. ISAC goes beyond RCC in pursuit of a deeper integration, where sensing and communications are more firmly integrated through a shared infrastructure and are delivered by the same waveform, optimizing both the sensing and communications performance. Current ISAC systems can be categorized into three classes, i.e., sensing-centric design, communication-centric design, and joint design and optimization, which are mainly based on design priorities and underlying signals and systems\cite{zhang2021}. On the one hand, integration gain can be achieved by sharing wireless resources for sensing and communications, thereby reducing duplication of infrastructure, devices, and transmissions. On the other hand, coordination gain can be obtained through mutual assistance of communications and sensing. 
\subsection{Massive MIMO LEO Satellite ISAC Systems}
The employment of massive MIMO can enhance the link budget for LEO satellites, facilitating the potential for wideband communication and sensing. ISAC presents a significant advantage for LEO satellites in efficiently using the limited onboard resources owing to the critical size, weight, and power constraints. In addition, the sensing information can be leveraged for beam training, beam tracking, and generic resource allocation \cite{isacoverview}. A massive MIMO LEO satellite ISAC system setup is illustrated in Fig. 1. The LEO satellite serves as a flying space BS to communicate with UTs while detecting targets or estimating quantities associated with targets in the sensing area of interest, e.g., angle-of-departure, velocity, etc.
\section{Recent Research Advances}
In this section, based on the analysis of the potential application scenarios, some recent research advances are presented.
\subsection{Multi-Satellite-Enabled ISAC}
Compared with the Global Navigation Satellite System operating at GEO and MEO, LEO satellite systems have several benefits, including 300 to 2400 times stronger signal
and threefold improvement in satellite geometry\cite{wang2021}. Moreover, driven by lower launching costs of LEO satellites, LEO satellite systems possess the capability to establish extensive global networks comprising numerous LEO satellites distributed across multiple orbital planes encircling the Earth, in which case a UT/target can be served by tens or even hundreds of LEO satellites simultaneously. With Multi-satellite cooperation in massive MIMO LEO satellite ISAC systems, the sensing and communications performance can be further improved.

The authors in \cite{wang2021} propose an integrated localization and communications framework to release the potential of LEO satellite networks. A novel cooperative beam hopping (BH) based solution is devised to adaptively adjust the physical beam layout between localization and communication beams in a time-division multiplexing manner. An example BH system in a 4-satellite scenario is illustrated in Fig. 2. As shown in Fig. 2(a), seamless coverage is realized with communication beams steered towards the coverage area centered at the LEO satellite nadir. The cooperative BH patterns for localization of Sat-1 and Sat-2 are depicted in Fig. 2(b) and Fig. 2(c), respectively. Simulation results obtained in \cite{wang2021} show neighboring satellite beams are steered to cover UTs/targets of interest, such that multiple good-quality signals can be measured to facilitate high-accuracy positioning with the proposed BH method. Besides,  with an increasing count of localization satellites, the enhancement of Cra$\acute{\text{m}}$er-Rao lower bound performance can be achieved significantly.
\subsection{RIS-Assisted Satellites for IoT Networks}
Along with the advent of ultra-massive connectivity among intelligent devices, the demand for ubiquitous connectivity in IoT networks is becoming more stringent than ever before. On the one hand, it is not easy to share sensing information of IoT networks in different areas just through terrestrial networks. On the other hand, sensing information obtained by IoT devices can assist communications, referred to as ISAC, further complicating the issue. These requirements necessitate novel wireless technologies to support 6G IoT networks.

The authors in \cite{tekbiyik2022} have proposed a new architecture involving the application of RIS units in massive MIMO LEO satellite systems for IoT networks. The high velocity of LEO satellites poses challenges to both communications and sensing. In this case, RIS units can be affixed onto the solar panels of LEO satellites, thereby fulfilling the imperative for steerable antennas capable of tracking. Specifically, as illustrated in Fig. 3(a) and Fig. 3(b), the incident wave can be scattered or beamformed to the UTs/targets by adjusting the phase shift of each reflecting element in accordance with the UTs/targets, and their propagation channels, such that the service quality for LEO satellite IoT networks can be enhanced\cite{tekbiyik2022}. Moreover, the UTs/targets without line-of-sight (LoS) connections can be communicated/sensed with RIS providing LoS links, as illustrated in Fig. 3(c). Simulation results in \cite{tekbiyik2022} show that the proposed architecture can yield achievable rates higher by a factor of $10^4$ when the incident wave is scattered and $10^5$ when the incident wave is beamformed for IoT networks.
\begin{figure*}[htbp]
	\centering
	\subfloat[]{\includegraphics[width=0.33\textwidth]{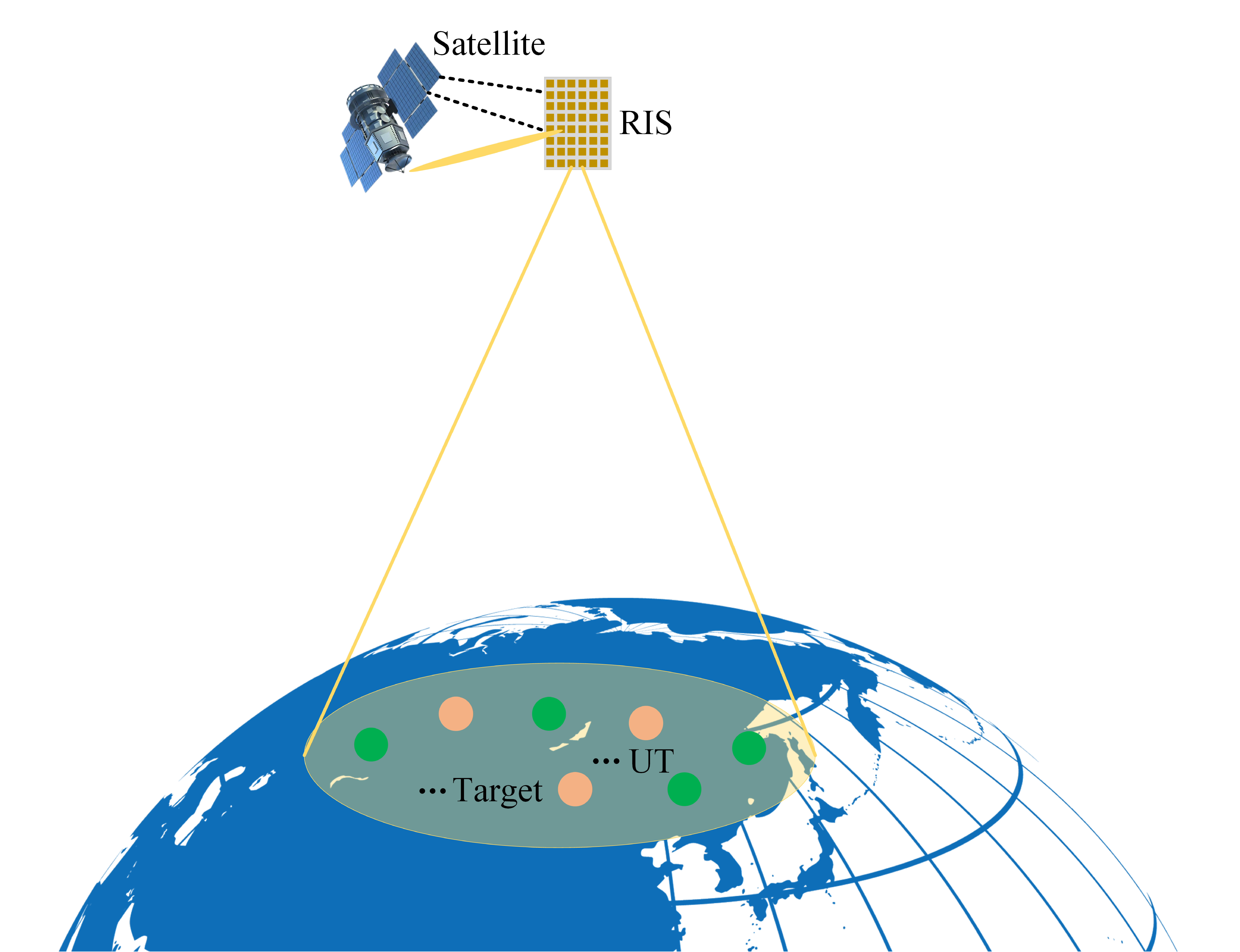}\label{}}
	\hfill
	\subfloat[]{\includegraphics[width=0.33\textwidth]{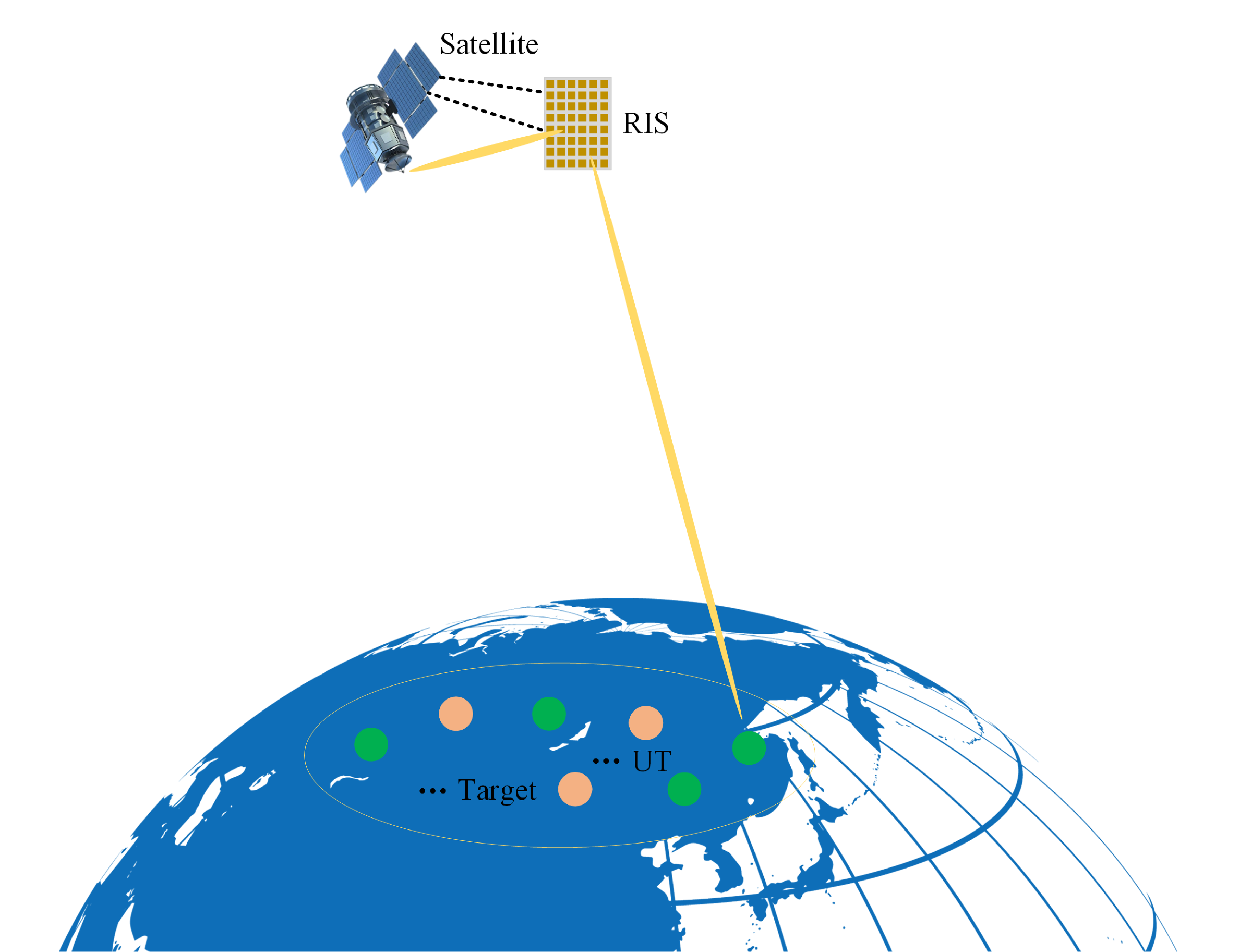}\label{}}
	\hfill
	\subfloat[]{\includegraphics[width=0.33\textwidth]{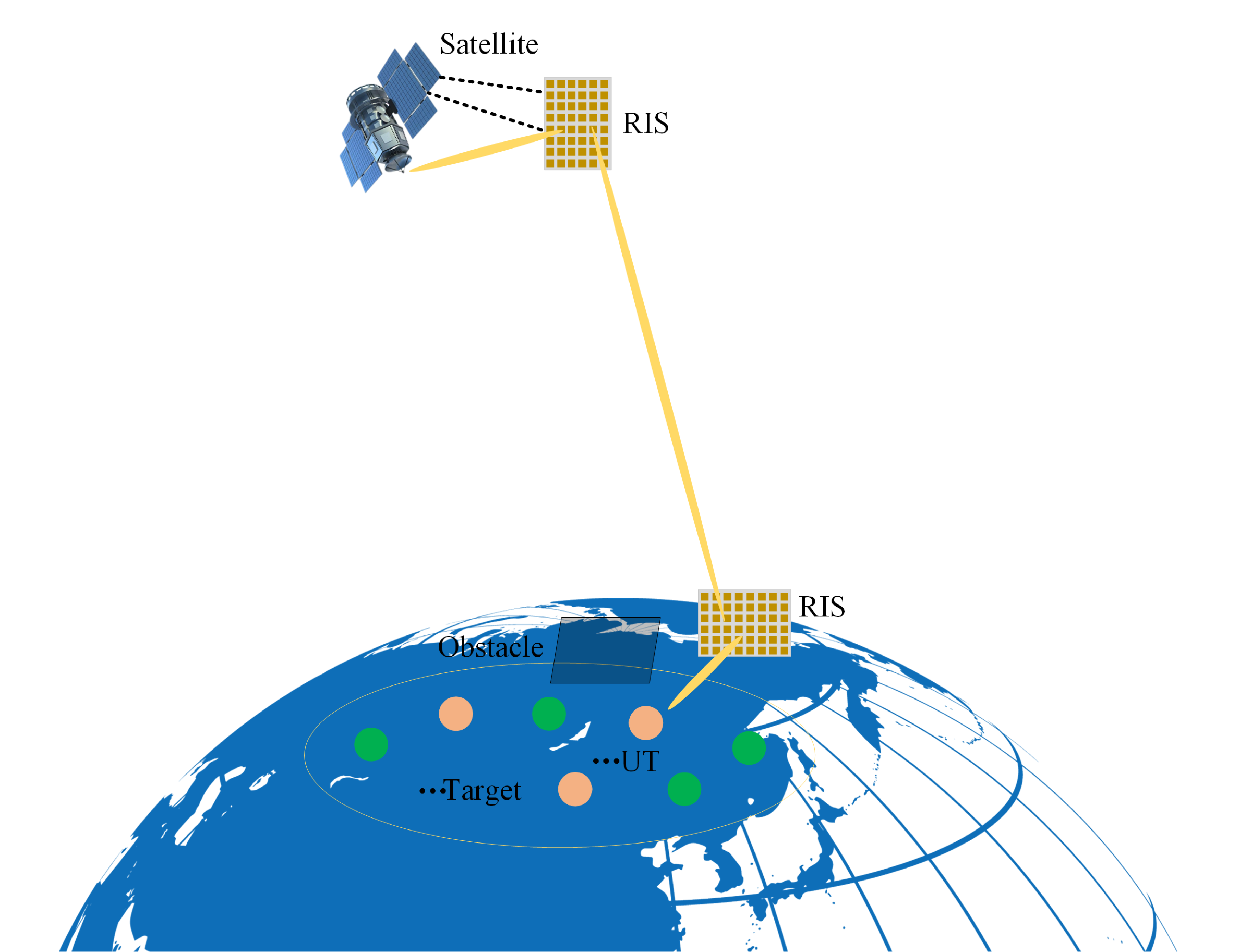}\label{}}
	\caption{(a) RIS units scatter the incident wave. (b) RIS units beamform the incident wave. (c) RIS units provide LoS links.}
	\label{fig3}
\end{figure*}
\section{Challenges and Key Technologies}
In this section, we will present some challenges in the considered systems and introduce several key enabling technologies.  
\subsection{Challenges}
\textit{1) Limited Payload Capabilities:} The limited payload capabilities of LEO satellites result in limited onboard resources, in which case the power consumption and hardware complexity should be given more consideration in massive MIMO LEO satellite ISAC systems. Furthermore, the massive MIMO structure in LEO satellite systems requires high hardware complexity if it is to be implemented in a fully digital way.

\textit{2) Non-Uniform Traffic Distribution:} Massive MIMO LEO satellite ISAC systems involve multi-dimensional resources such as time, frequency, space, and space, which puts forward high requirements for resource allocation schemes. However, due to the non-uniform traffic distribution over the coverage area, it is intractable to make the optimal match between resources and different types of services.

\textit{3) Synchronization Tracking and Mobility Management:} LEO satellites move at a high speed of 5 -- 10 km/s, resulting in a series of performance challenges, including time synchronization tracking, frequency synchronization tracking caused by large Doppler shifts. In addition, when addressing mobility management, it is crucial to account for the limited temporal visibility window on Earth\cite{visibilitywindow}.

\textit{4) Beam Squint Effects:} A large number of antennas and wide bandwidth might result in frequency-dependent array responses and severe beam squint effects, where the beams would steer towards different directions from different subcarriers, making the energy of part subcarriers deviate from the locations of UTs/targets\cite{you2022oct}.
\subsection{Key Technologies}
\textit{1) Software Defined Payloads and Networks:} Massive MIMO LEO satellite ISAC systems require onboard flexible load adjustment capability. Software defined payloads (SDP) provide a feasible solution by configuring LEO satellite functionalities on demand. Through programmable hardware and virtualization of network functions, LEO satellite payloads can be reconstructed for various missions\cite{zhangma2021}. In addition, software defined networks (SDN) virtualize hardware and services previously performed by dedicated hardware, which minimizes the hardware space taken up\cite{SDN}. Software defined payloads and networks can be exploited in massive MIMO LEO satellite systems to maximize hardware integration and shorten mission response time, thus improving the sensing and communications performance simultaneously.

\textit{2) Hybrid Beamforming:} Hybrid beamforming is capable of reducing power consumption and computational complexity while ensuring system performance, which depends heavily on accurate channel state information (CSI). Nonetheless, due to the large Doppler shifts and high propagation delay, it is challenging to obtain accurate instantaneous CSI at the LEO satellite's side. In comparison to instantaneous CSI, statistical CSI changes over large time scales, which can be acquired with sufficiently high accuracy. Recently, some efforts have been endeavored to exploit hybrid beamforming based on statistical CSI. The authors in \cite{you2022oct} design the transmitter of massive MIMO LEO satellite ISAC systems that simultaneously implement communications and sensing functionalities based on statistical CSI. A weighted sum method for the problem with transformed objectives to trade-off between the performance of communications and sensing is employed, where a weighting coefficient is introduced to adjust the weight between the communication and sensing modules. Besides, an efficient algorithmic approach is proposed to alleviate the beam squint effects. The energy efficiency performance of the communication module versus the power budget under different weighting coefficients and the corresponding sensing beampattern with a particular coefficient is shown in Fig. 4(a) and Fig. 4(b), respectively. Simulation results demonstrate that the beams can be steered towards the targets of interest while guaranteeing the performance of energy efficiency.
\begin{figure}[htbp]
	\centering
	\subfloat[]{\includegraphics[width=0.5\textwidth]{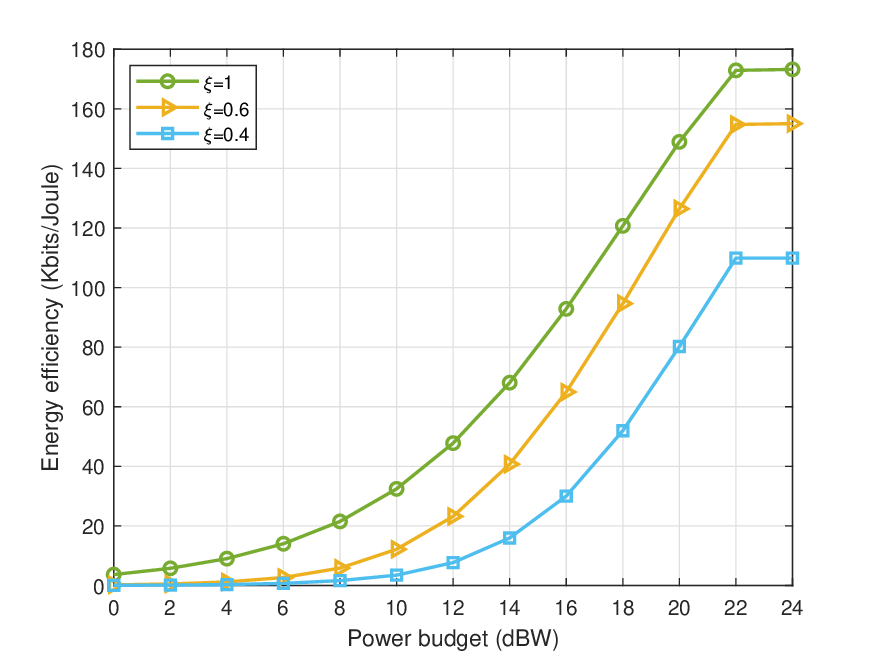}\label{}}
	\hfill
	\subfloat[]{\includegraphics[width=0.5\textwidth]{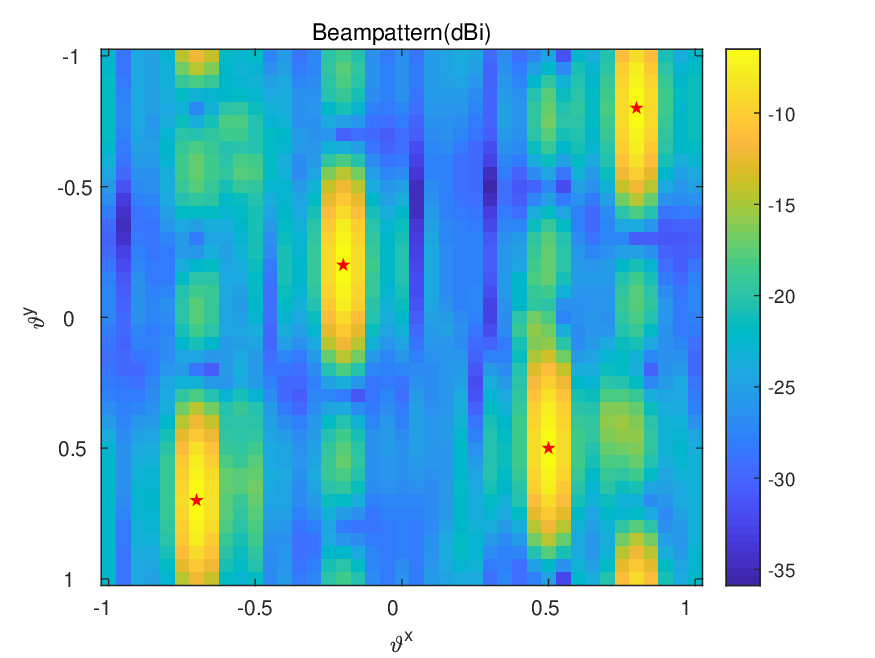}\label{}}
	\caption{(a) Energy efficiency performance versus power budget $P$ with 256 antennas under different weighting coefficient $\xi$. (b) Beampattern versus space angles with $\xi$ = 0.6. The red pentagrams locate the position of the targets.}
	\label{fig4}
\end{figure}

\textit{3) Edge Intelligence:} Edge intelligence is considered an enabling technology for 6G networks, which can move the computation-intensive work from the centralized cloud to distributed BSs at the wireless network edge to make efficient use of large amounts of data generated at various devices\cite{edge_intelligence}. Particularly, to achieve high-capacity communications and flexible sensing in the considered massive MIMO LEO satellite ISAC systems, the large amount of raw data generated at distributed wireless transceivers demanded to be appropriately processed with edge intelligence in an energy-efficient and swift manner owing to the limited available onboard resources and high velocity of LEO satellites.

\section{Open Issues and Research Opportunities}
In this section, we will briefly discuss some open issues and promising research directions in massive MIMO LEO satellite ISAC systems.
\subsection{Novel Antenna Architecture Design}
Extremely large-scale MIMO technology has received considerable attention in 6G networks, which will further benefit LEO satellite ISAC systems. However, the significant surge in the number of antennas will lead to huge physical size and high hardware complexity, where novel antenna architecture needs to be well designed to alleviate these drawbacks. As a brand-new antenna paradigm, holographic metasurface antennas (HMAs) can facilitate the layout of extremely large-scale antenna arrays since the metamaterial elements can be distributed with sub-wavelength intervals. Besides, HMA-based transceivers need much fewer RF chains than conventional transceivers, such that the power consumption can be significantly reduced. In addition, lens antenna arrays have inherent advantages in reducing hardware complexity. The communication capacity and sensing accuracy can both be improved by utilizing the energy focusing property of extremely large-scale antenna arrays.
\subsection{Space-Air-Ground-Sea Integrated Network (SAGSIN) Security With ISAC}
As mentioned in Sec. \uppercase\expandafter{\romannumeral1}, 6G will harmonize satellite, aerial, terrestrial, and marine networks to achieve three-dimensional global coverage. Operating ISAC in massive MIMO LEO satellite systems enables information exchange among different networks, a mission convenient and efficient, which further improves the integration gain in SAGSIN. However, security issues emerging in integrated systems are largely overlooked in recent studies. First, the routing node lacks physical protection in SAGSIN, and its mobility and flexibility constrain the application of complex cryptographic algorithms. This vulnerability increases the risk of enemy control or capture. Besides, SAGSIN presents increased exposure of channels to adversaries compared to other typical networks, and the simplicity of link establishment within the network further heightens security risks. In addition, due to the shared waveform for sensing and communications, critical information could be leaked to the sensing targets. Toward this end, in-depth investigation is highly demanded to tackle the security issues in integrated systems.
\subsection{Novel Waveform Design}
To fully reap the benefits of the envisioned massive MIMO LEO satellite ISAC system, a fundamental issue lies in the dual-functional waveform design. As for communication waveform design, several key parameters, including peak-to-average power ratio error performance and spectral efficiency, should be considered. Concerning sensing waveform design, it is critical to take detection probability and recognition accuracy into account\cite{isacoverview}. However, traditional orthogonal frequency division multiplexing waveform suffers from a performance loss due to the large Doppler shifts in LEO satellite systems. Recently, the orthogonal time frequency space (OTFS) modulation has been considered as a candidate technology for LEO satellite systems to efficiently overcome severe Doppler effects.
\section{Conclusion}
Operating the ISAC in the massive MIMO LEO satellite systems has great potential to provide wide coverage for wireless communications and sensing. However, some technical issues demand to be addressed in the considered systems. This paper highlighted several characteristics of LEO satellite communication systems and briefly offered an overview of ISAC and the considered massive MIMO LEO satellite ISAC systems. The relevant research advances, including multi-satellite-enabled ISAC and RIS-assisted satellites for IoT networks, were presented. Besides, we identified some challenges in the considered systems, like limited payload capabilities, non-uniform traffic distribution, synchronization tracking, mobility management, etc. Some key enabling technologies, including SDP, SDN, hybrid beamforming, and edge intelligence, were discussed. Finally, we pointed out several open issues and promising research directions, such as novel antenna architecture design, SAGSIN security with ISAC, and novel waveform design. We believe that this paper would inspire more innovative ideas for this important research topic for the coming 6G networks.
\bibliographystyle{IEEEtran}  
\bibliography{reference}
\end{document}